\documentclass[longbibliography,prb,twocolumn]{revtex4-1}
\usepackage{graphicx}
\usepackage{dcolumn}
\usepackage{booktabs,bm,color,braket,amsmath}
\usepackage{color}
\usepackage[dvipsnames]{xcolor}

\definecolor{DarkRed}{RGB}{100,0,0}

\usepackage{microtype} 
\usepackage[hidelinks,colorlinks,linkcolor=blue,
citecolor=blue,urlcolor=blue]{hyperref}

\usepackage[normalem]{ulem}

\begin{document}

\title{Chirality induced Giant Unidirectional Magnetoresistance in Twisted Bilayer Graphene}

\author{Yizhou Liu, Tobias Holder, Binghai Yan}
\email{binghai.yan@weizmann.ac.il}

\affiliation{
Department of Condensed Matter Physics,Weizmann Institute of Science, Rehovot 76100, Israel
}

\begin{abstract}
\textbf{Twisted bilayer graphene (TBG) exhibits fascinating correlation-driven phenomena like the superconductivity and Mott insulating state, with flat bands and a chiral lattice structure. We find by quantum transport calculations that the chirality leads to a giant unidirectional magnetoresistance (UMR) in TBG, where the unidirectionality refers to the resistance change under the reversal of the direction of current or magnetic field. We point out that flat bands significantly enhance this effect. The UMR increases quickly upon reducing the twist angle, and reaches about 20\% for an angle of 1.5$^\circ$ in a 10 T in-plane magnetic field. We propose the band structure topology (asymmetry), which leads to a direction-sensitive mean free path, as a useful way to anticipate the UMR effect. The UMR provides a probe for chirality and band flatness in the twisted bilayers. }
\end{abstract}

\maketitle

\section{Introduction.}

Twisted bilayer graphene (TBG) with a small interlayer angle forms a chiral moi\'re superlattice, whose superlattice minibands exhibit greatly reduced velocity~\cite{Bistritzer2011,Castro2007,Suarez2010}. The flat bands at the magic angle near $1^\circ$ provide an ideal platform to study correlation-driven phenomena~\cite{Cao2018Mott,Cao2018SC,Goldhaber2019,Young2019,Efetov2019,Wang2019,Andrei2019,Nadj-Perge2019,Pasupathy2019,Yazdani2019,Wang2020,Ilani2020,Yazdani2020}, for example, unconventional superconductivity and the Mott insulating state. TBG has also motivated the design of twisted structures of other van der Waals materials~\cite{MacDonald2019tmd,Yu2019,An2019WSe2,Zhang2019tmd,Park2020WSe2,Sarma2020,Merkl2020,Lian2020,WangL2020,Regan2020,Tang2020,Naik2018}. In contrast, chirality-induced transport phenomena are less explored thus far~\cite{liu2020anomalous,He2020,hu2020nonlinear,Huang2020WSe2NLAHE}. 


The chirality and flat bands motivate us to investigate the electric magnetochiral anisotropy (EMCA)~\cite{Rikken2001}, which is characterized by the unidirectional magnetoresistance (UMR) in twisted bilayer systems. In an ordinary material, the resistance remains the same upon reversing the direction of the current or the magnetic field. In a chiral system, however, the magnetoresistance (MR) is nonreciprocal, and the chirality determines the preferred direction of the UMR. We note that the EMCA is different from the known anisotropic magnetoresistance (AMR), which remains the same when reversing the magnetic field direction. The EMCA was observed for example in Bismuth helices~\cite{Rikken2001}, DNA~\cite{Xie2011,Liu2020chiral}, and other quantum materials \cite[and references therein]{Tokura2018}, in which the relative amplitude of UMR was usually several percent. Very recent experiments based on the quantum spin Hall edges~\cite{Zhao2020WTe2} and quantum anomalous Hall edges~\cite{Yasuda2020QAHE} reported giant UMR (according to the definition of Eq.~\eqref{Eq:ratio}) of 33\% and 13\% , respectively. This large UMR promises applications in optoelectronic and spintronic devices. The UMR represents a current rectification, which can generate a dc current in an ac electric field and can be controlled by the magnetic field. This rectification effect can be used for photon-detection or energy-harvesting from long-wavelength light. The dependence of the UMR on the direction and amplitude of the magnetic field provides a sensor for the vector magnetic field.

Different from ordinary transport experiments, the UMR measurement probes the symmetry breaking and dissipative process such as magnetic excitations~\cite{Yokouchi2017}. It is commonly attributed to the inelastic scattering by magnons~\cite{Yasuda2016} or spin clusters~\cite{Ishizuka2020}. However, it is unclear how the intrinsic band structure acts in UMR. Especially for the twisted bilayers, it is an interesting question which role flat bands play for the UMR. 

In this work, we find that flat bands generally induce a large UMR for TBG through the orbital effect in an in-plane magnetic field. Our quantum transport calculations demonstrate an increasing UMR when reducing the twist angle down to the magic angle. The two-terminal MR non-reciprocity is determined by two factors: different electron velocities along counter-propagating directions, and the dephasing in the transport. With these two ingredients, we derive a formula to estimate the UMR from the band structure, which agrees quantitatively with our calculations within the Landauer-B\"uttiker formalism. Here, the magnetic field induces direction-dependent Fermi velocities in the band structure and thus leads to a direction-dependent mean free path, resulting in the UMR effect.  The UMR is inversely proportional to the Fermi velocity and thus strongly enhanced by flat bands. It can reach about 20\% for smaller twist angle 1.5$^\circ$ in an in-plane magnetic field of 10 T. Because the UMR effect originates in the symmetry-breaking (both time-reversal- and inversion-breaking), it can exist even if strong electron-electron interactions renormalize the band structure. 


\section{Methods}

\begin{figure}
    \centering
    \includegraphics[width=\linewidth]{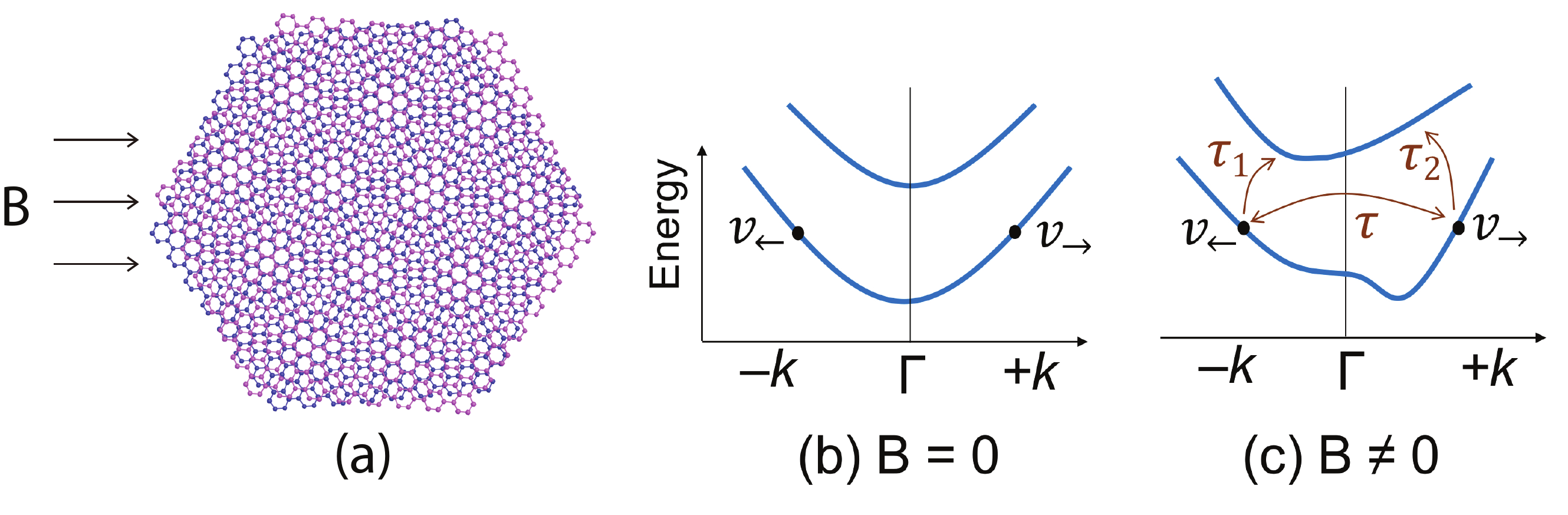}
    \caption{\label{fig1} \textbf{The chiral structure of TBG and illustration of the band structure asymmetry.} (a) The lattice structure of a TBG. The red and blue spheres represent carbon atoms in the top and bottom layers, respectively. The TBG belongs to the chiral point group $D_6$. (b-c) Schematics of the band symmetry-breaking of a chiral system induced by the magnetic field ($\mathbf{B}$). (b) Without the magnetic field, the left- and right-moving electrons have the same velocity ($v_{\rightarrow}=v_{\leftarrow}$) at generic Fermi energy. Corresponding left- and right-moving conductance are equal, $G_{\rightarrow}=G_{\leftarrow}$. (c) If finite $\mathbf{B}$ breaks the time-reversal symmetry, the velocity balance is violated($v_{\rightarrow}>v_{\leftarrow}$). Thus, the conductance becomes nonreciprocal, i.e., $G_{\rightarrow}(\mathbf{B}) > G_{\leftarrow}(\mathbf{B})$, according to Eq.~\ref{Eq2-G}. The red arrows indicate the principle of how micro-nonreversibility happens. Based on Fermi's Golden rule the ordinary relaxation rate $\tau$ between each other enforces the microreversibility i.e. $\tau_\leftarrow =\tau_\rightarrow$. However, the nonreciprocity can exist when second relaxation channels ($\tau_1$ and $\tau_2$) are taken into account. 
  }
\end{figure}

The TBG belongs to the chiral point group $D_6$, which exhibits no inversion or mirror symmetry. In a system with either inversion symmetry {(or the two-fold rotation around the $z$-axis for a 2D material)} or time-reversal symmetry, the left and right moving electrons have the same velocity at the Fermi surface. If both inversion and time-reversal symmetry are broken, the left and right movers exhibit different Fermi velocities, as illustrated in Fig.~\ref{fig1}. This is the case for chiral TBG in an in-plane magnetic field. The velocity imbalance leads to the UMR, as we will discuss in the following. In contrast, the out-of-plane field, which preserves the {two-fold rotation around $z$}, requires further symmetry reductions to break the band structure symmetry of TBG.

We adopt the tight-binding Hamiltonian to describe the band structure of TBG outlined in Ref.~\cite{Moon2013}. This tight-binding model includes all atomic sites of two layers and is compatible with the introduction of an in-plane magnetic field. We consider the orbital effect of the magnetic field [$\mathbf{B}=(B_x, B_y, 0)$] and ignore the weak spin splitting and negligible spin-orbit interaction. Suppose a two-terminal device where the current flows along the $x$ direction. We choose the gauge $\mathbf{A} = (B_y z,-B_x z,0)$ so that the lattice translational symmetry along $x$ (the charge current direction) is still preserved. The modification of the hopping integrals by the magnetic field is taken into account via a Peierls substitution~\cite{Peierls1933,Xiao2010RMP}. Because it promotes the inversion symmetry and loses the chiral character, we do not employ the Bistritzer-MacDonald~\cite{Bistritzer2011} continuum model for TBG even though it can easily include an in-plane magnetic field~\cite{Kwan2020,Roy2013}.
 
\begin{figure}
    \centering
    \includegraphics[width=\linewidth]{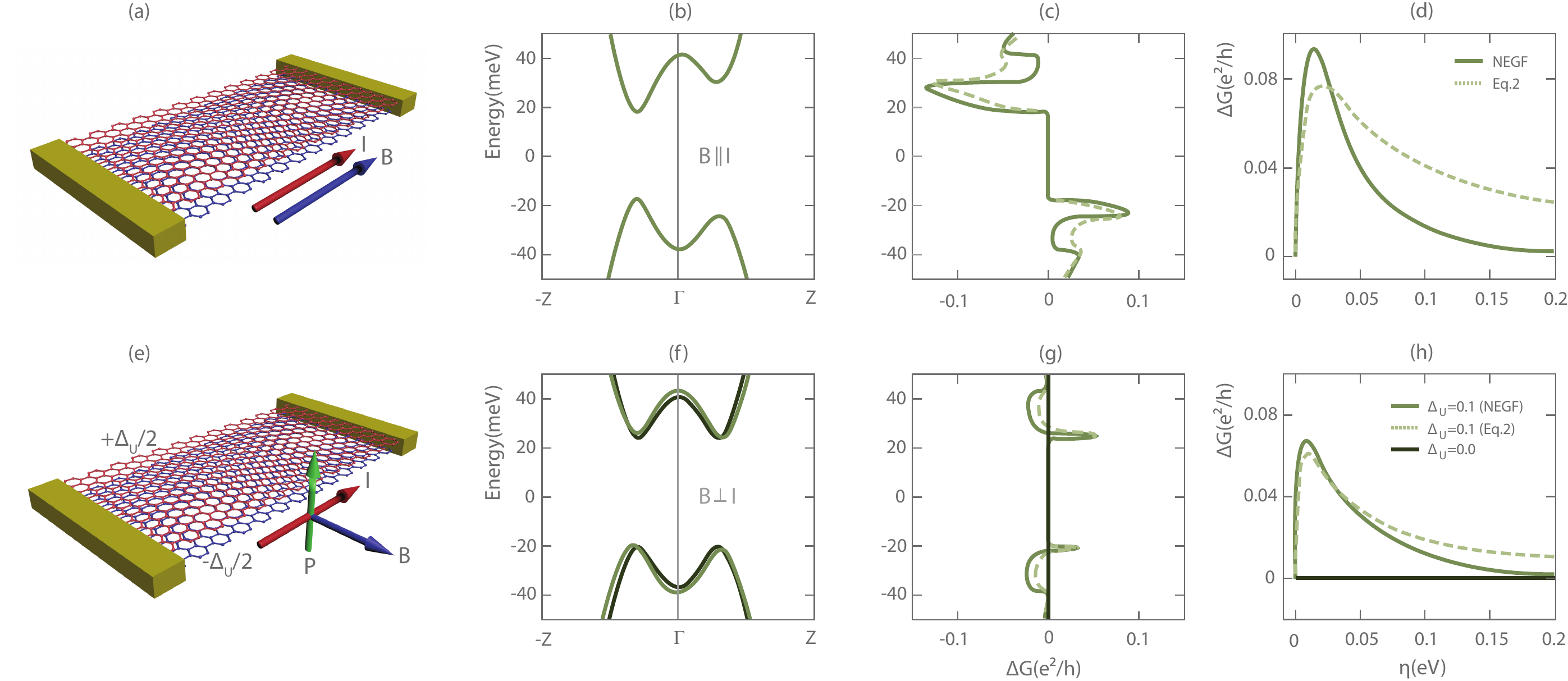}
    \caption{\label{fig2} \textbf{The band structure and conductance of TGB with twisted angle 3.15$^\circ$}. (a) Schematic of the transport device of TBG under in-plane magnetic field $\mathbf{B||I}$. (b) The band structure for the nanoribbon (the device region in a) with B= 50 T. (c) The direction-induced conductance change $\Delta G = G_\rightarrow - G_\leftarrow$ at different Fermi energies. Results from the NEGF method and Eq. 2 agree very well. The conductance is calculated with a constant dephasing parameter $\eta = 0.01$ eV.  (d) The peak value of $\Delta G$ for the valence band top depends non monotonically on the dephasing parameter $\eta$. The NEGF and Eq. 2 provides similar results, especially in the small $\eta$ regime. (e) The device for $\mathbf{B \perp I \perp P}$. $\mathbf{P}$ is the charge dipole induced by the parameter $\Delta_U$. (f) Corresponding band structures for $\Delta_U=0, 0.1$ eV. The applied magnetic field is B = 50 T. (g) Corresponding $\Delta G$ at different Fermi energies. (h) The $\eta$-dependence of $\Delta G$. Solid curves are calculated by NEGF and the dashed one is evaluated from Eq. 2. }
\end{figure}

To evaluate the UMR, we calculate the total conductance of a two-terminal device with an in-plane magnetic field by the nonequilibrium Green's function (NEGF) method~\cite{Datta2012}. The whole system is divided into three parts: two semi-infinite reservoirs acting as electron leads, and a scattering region connecting them. We set these three parts as the same TBG ribbons along the $x$ direction, which preserve the $D_6$ point group symmetry (cf. Fig.~\ref{fig2}{\textcolor{blue}{a}}). The conductance from lead $q$ to $p$ is $G_{p\leftarrow q} = \textrm{Tr}(\Gamma_p \mathcal{G} \Gamma_q \mathcal{G}^\dag)$ where $\Gamma_{p(q)}=i(\Sigma_{p(q)} - \Sigma^\dag_{p(q)})$ is the anti-Hermitian part of the retarded self energy matrix $\Sigma_{p(q)}$ of lead $p(q)$; $\mathcal{G} = (E + i\eta)I - H - \Sigma$ is the Green function of scattering region where $I$ is the identity matrix, $H$ is the Hamiltonian, $\Sigma$ is the total self-energy from coupling to the leads, and $\eta$ is the dephasing parameter~\cite{Buttiker1986b}. If $\eta=0$ one can prove the reciprocal relation $G_{p\leftarrow q}=G_{q\leftarrow p}$, regardless of the presence of a magnetic field~\cite{SM}. This amounts to the fact that in coherent transport the two-terminal conductance depends only on the total channel number but not on the velocity~\cite{Buttiker1986b}. However, a finite $\eta$ spoils this reciprocity, as we  discuss now.

\section{Results}

We first demonstrate the UMR effect by a phenomenological model. Suppose a two-terminal device with the scattering region of length $L$, whose conductance can be evaluated as \cite{Datta2012},
\begin{equation}\label{EquationG}
    G_0 = G_B \frac{\lambda}{\lambda + L},
\end{equation}
where $G_B$ is the ballistic conductance and $\lambda$ is the phase-coherent length or the mean free path. Equation~\eqref{EquationG} connects the ballistic and diffusive limits, which correspond to $\lambda \gg L$ and \textbf{$\lambda \ll L$  }, respectively. For a finite dephasing parameter $\eta$, $\lambda_{\rightarrow/\leftarrow} = v_{\rightarrow/\leftarrow} \hbar/\eta $ where $v_{\rightarrow/\leftarrow}$ is the right/left-moving Fermi velocity. We generalize Eq.~\eqref{EquationG} to evaluate the conductance ($G_\rightarrow$) and resistance ($R_\rightarrow $) from the left to the right as
\begin{align}
\label{Eq2-G}
G_\rightarrow & = G_B \left(1 + \frac{\eta L }{\hbar v_\rightarrow}\right)^{-1},\\
\label{Eq2-R}
R_\rightarrow & = R_B \left(1 + \frac{\eta L }{\hbar v_\rightarrow}\right),
\end{align}
where $G_B$ is the ballistic conductance and $R_B \equiv 1/G_B$. 
At finite $\eta$, the directional conductance or resistance depends on the corresponding Fermi velocity. Equations~(\ref{Eq2-G},\ref{Eq2-R}) give an intuitive explanation of a mechanism for the UMR. They also provide an alternative way to quantitatively estimate the unidirectional conductance from the band structure of a mesoscopic system.

The UMR can further be justified in a semiclassical picture where the left- and right-moving electrons scatter mostly between each other within the same band. But there is additionally some weak scattering into states provided by a nearby band. This side-channel of relaxation leads to an additional resistance with a coefficient that depends on the DOS ($\tau_{\leftarrow} \neq \tau_{\rightarrow}$), yielding Eq.~\eqref{Eq2-G} (see more in the SM~\cite{SM}). Therefore, if $v_\rightarrow > v_\leftarrow$ (Fig.~\ref{fig1}{\textcolor{blue}{c}}), this yields nonreciprocal transport with $G_\rightarrow > G_\leftarrow$ or $R_\rightarrow <  R_\leftarrow$.  We assume that the dephasing in both directions is equal. In addition, Ref.~\cite{ideue2017bulk} also discussed the band structure asymmetry to induce the nonreciprocal transport in the diffusive regime described by the second-order Boltzmann equation.

Furthermore, we find that flat bands can generally enhance the UMR because
\begin{align}
\frac{\Delta R}{R}  &= \frac{R_\rightarrow - R_\leftarrow}{R_\rightarrow + R_\leftarrow} = \frac{-\Delta G}{G}  \label{Eq:ratio} \\
&= \frac{v_\leftarrow-v_\rightarrow}{\frac{2\hbar v_\leftarrow v_\rightarrow}{\eta L}+v_\rightarrow+v_\leftarrow} 
\label{Eq:UMR1}\\
&\stackrel{\eta L \to \infty}{=} \frac{v_\leftarrow-v_\rightarrow}{v_\rightarrow+v_\leftarrow} \propto (v_\leftarrow-v_\rightarrow) \cdot \mathrm{DOS}.
\label{Eq:UMR2}
\end{align}
Equation~\eqref{Eq:UMR2} indicates that the upper limit of the UMR is an intrinsic quantity determined by the field-modified band structure, non-withstanding the fact that it requires finite dephasing/dissipation. The term $1/(v_\rightarrow+v_\leftarrow)$ in Eq.~\eqref{Eq:UMR2} represents the density of states (DOS), and thus the amplitude of the UMR is approximately proportional to the DOS. Additionally, the velocity imbalance requires breaking of both the inversion symmetry and time-reversal symmetry. The time-reversal breaking can be realized by an external magnetic field or spontaneous magnetization in TBG~\cite{Goldhaber2019,Young2019}. It should be noted that Eqs.~\eqref{Eq:UMR1} and \eqref{Eq:UMR2} are only accurate for the single-mode case at the Fermi energy. The $\eta L$ dependence of the UMR can be nonmonotonic in the many-band case (see the SM for more details). {On the other hand, we do not expect qualitative changes to the phenomenology in the presence of strong electron-electron interactions, which to a large extent preserve momenta and thus influence transport only indirectly.}

\textit{Magnetochiral effect of TBG. ---}In the presence of an in-plane magnetic field ($\mathbf{B}$), the symmetry-allowed corrections to the resistance/conductance of the current $\mathbf{I}$ are in the form of $\mathbf{I \cdot B}$ and $\mathbf{I \times B}$. They satisfy the global Onsager's reciprocal relation~\cite{Buttiker1986}, i.e., $R(\mathbf{I,B})=R(\mathbf{-I,-B})$. The resistance in EMCA is usually written as~\cite{Rikken2001,Rikken2005},
\begin{equation}
\label{Eq:EMCA}
R(\mathbf{I,B})=R_0 + \chi^{L,R}\mathbf{I \cdot B} + \gamma \mathbf{(I \times B)\cdot P}{ + \beta \mathbf{B}^2}, \end{equation}
where $\chi^{L,R}$ represents the chirality with $\chi^{L}=-\chi^{R}$, $\gamma$ add{and $\beta$} {are} constants, and $\mathbf{P}$ the charge polarization. {The last $B^2$ term represents the ordinary MR due to the Lorentz force, which is reciprocal.}

Equation~\eqref{Eq:EMCA} indicates the nonlinear I--V dependence and the second-harmonic resistance induced by an AC electric field. We point out that the dephasing term $\eta$ in NEGF calculations or Eqs.~\eqref{Eq2-G} and \eqref{Eq2-R} includes both elastic and inelastic scattering. The inelastic scattering leads to the nonlinear effects.

In TBG, the chirality refers to the twisted direction, left- or right-handed. In the device, the h-BN substrate induces the sublattice symmetry-breaking~\cite{Jung2015,Hunt2013,Goldhaber2016,Senthil2019,Bultinck2020,Kim2018}, reducing the $D_6$ symmetry to $C_3$ and leading to an out-of-plane dipole. The in-plane strain can further break the $C_3$ rotation to $C_1$~\cite{Pasupathy2019,Nadj-Perge2019} and induces an in-plane dipole. For the $D_6$ symmetry, the chiral term $\mathbf{I \cdot B}$ contributes to the magneto-transport. For the case of $C_3$ symmetry, the additional dipole term $\mathbf{(I \times B)\cdot P}$ emerges. We will focus on these two terms in the following and later discuss the $C_1$ case with the in-plane dipole. It is convenient to calculate the conductance, compared to the resistance, with the NEGF method. Thus, in the following, we evaluate the unidirectional magneto-conductance (UMC).


First, we take TBG with the twist angle $\theta=3.15 ^\circ$ to demonstrate the UMC for both cases, $\mathbf{B||I}$ and $\mathbf{B \perp I}$. We calculate the conductance change $\Delta G = G_\rightarrow (\mathbf{B})-G_\leftarrow (\mathbf{B})$ through a TBG nanoribbon by NEGF. The outer section of the nanoribbon is used as two leads in the device, as shown in Fig.~\ref{fig2}{\textcolor{blue}{a}}. The applied magnetic field $\mathbf{B||I}$ leads to a phase in the interlayer hopping and breaks the velocity balance at opposite $k$, a consequence of time-reversal breaking. This $k\to-k$ symmetry breaking is shown in the band structure in Fig.~\ref{fig2}{\textcolor{blue}{b}}. To demonstrate the band structure change and $\Delta G$ clearly, we apply a large magnetic field 50 T for this twist angle.


Subsequently, UMC emerges with nonzero $\Delta G$. The peaks of $\Delta G$ originate from band edges where the velocity approaches zero and the DOS exhibits a peak, consistent with the prediction of Eq.~\eqref{Eq2-G}. We point out that $\Delta G$ depends on the dephasing parameter $\eta$. If $\eta=0$, which represents fully coherent transport, we obtain $\Delta G = 0$ in the NEGF calculations and also according to Eq.~\eqref{Eq2-G}. The opposite limit $\eta\rightarrow \infty$ also yields $\Delta G \rightarrow 0$, because electrons do not remember their group velocity in the diffusive limit without any coherence. A large $\Delta G$ appears in the intermediate region (see Figs.~\ref{fig2}{\textcolor{blue}{c-d}}), where the finite $\eta$ violates the charge conservation in the two-terminal conductance because it effectively introduces a non-Hermitian term $i\eta$ into the Hamiltonian thus breaking the probability conservation of Schr\"odinger equation~\cite{Buttiker1986b}. Therefore, the existence of a two-terminal UMC/UMR requires both dephasing and partial coherence, which can be induced by the inelastic scattering.

Next, we rotate the magnetic field as $\mathbf{B \perp I}$. We further induce an electric dipole $\mathbf{P}$ out of the plane by adding a potential difference  $\Delta_U$ between both layers (see Fig.~\ref{fig2}{\textcolor{blue}{e}}). Then, a UMC emerges due to the $\mathbf{(B \times I)\cdot P}$ term in Eq.~\eqref{Eq:EMCA}. The value of $\Delta G$ is approximately proportional to the potential $\Delta_U$ and also relies on a finite $\eta$. Similar to the case $\mathbf{B || I}$, the band edges constitute the peaks in $\Delta G$.

We stress that Eq.~\eqref{Eq2-G} agrees quantitatively with the NEGF calculations especially in the small $\eta$ regime, as demonstrated in Figs. \ref{fig2}{\textcolor{blue}{c}}, \ref{fig2}{\textcolor{blue}{d}} and \ref{fig2}{\textcolor{blue}{g}}. The deviation in the regime of large $\eta$ is likely because Eq.~\eqref{Eq2-G} neglects interference effects~\cite{Baranger1991}. Such effects come into play once further relaxation processes are taken into account for the electrons that are already relaxed via the side-channel. However, these processes are higher-order not only in terms of $\eta$ but also microscopically in terms of processes that lead to a finite $\eta$. Therefore, to evaluate the UMC for smaller twist angles where the NEGF demands much more computational time, we will use Eq.~\eqref{Eq2-G} without sacrificing much accuracy.


The nanoribbon configuration indicates that the flatness of the band edges leads to a large UMC. However, the finite width (e.g., one moir\'e unit in Fig. \ref{fig2}) and the existence of edge states implies that the ribbon configuration can exhibit different transport behavior from the extended 2D bilayer. For this reason, we investigate in the following a sheet of TBG with infinite width for several twist angles. To this end, the UMC is extracted by summing the 1D conductance (Eq.~\eqref{Eq2-G}) over the transverse momenta. 

\begin{figure}
    \centering
    \includegraphics[width=0.95\linewidth]{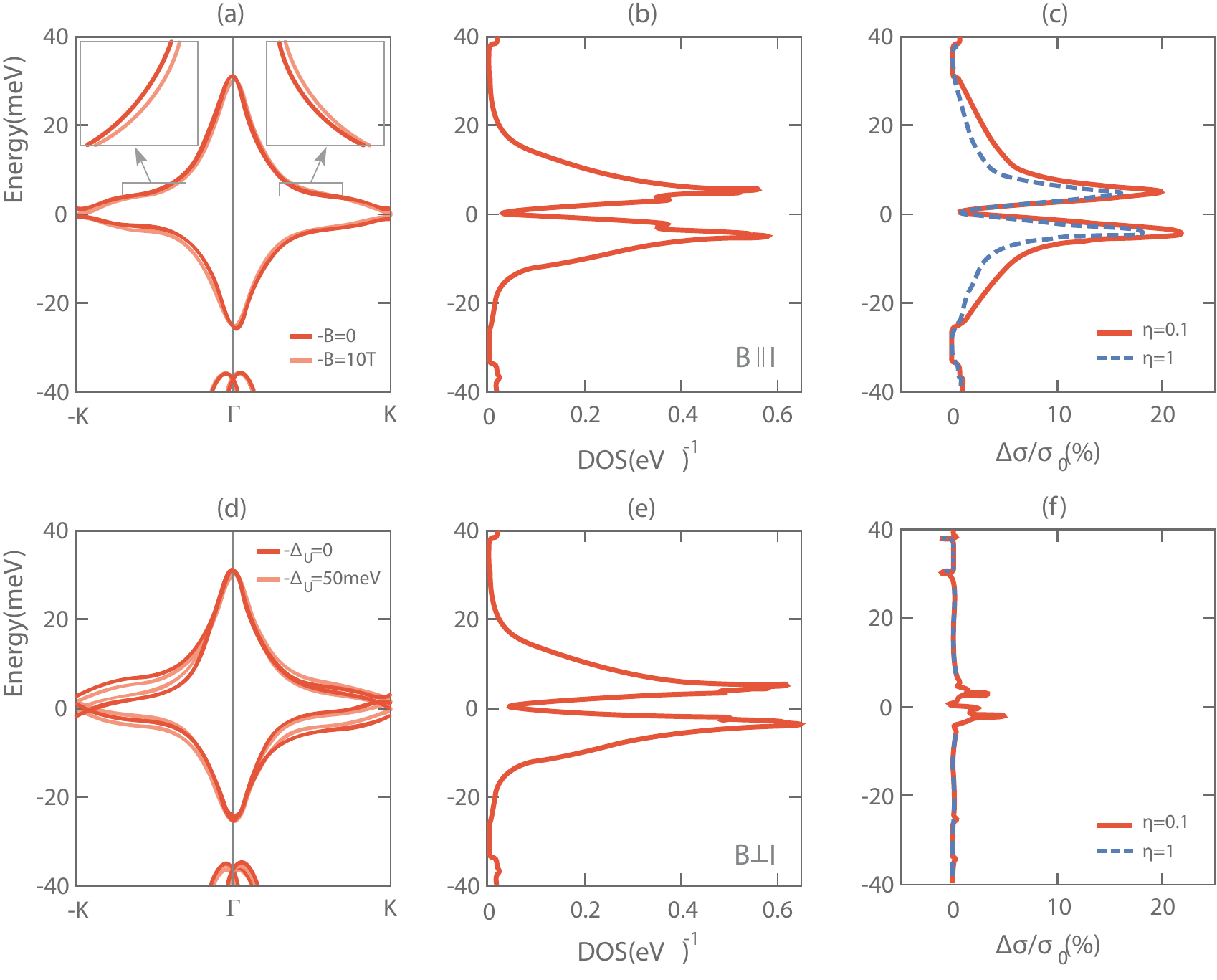}
    \caption{\label{fig3}\textbf{Band structure and UMR of the TBG with $\theta=1.5^\circ$ for $\mathbf{I} \parallel \mathbf{B}$ [(a)-(c)] and $\mathbf{I} \perp \mathbf{B}$ [(d)-(f)].} (a)\&(d) The band structure along high-symmetry lines. For $\mathbf{I} \parallel \mathbf{B}$ black (red) lines represent the results for B = 10 (B = 0) T; for $\mathbf{I} \perp \mathbf{B}$ they represent results for $\Delta_U=0$ ($\Delta_U=50$) meV with fixed magnetic field $B=10$ T. The Fermi energy is set to zero, i.e., the charge neutral point. (b)\&(e) Density of states (DOS) under the magnetic field. (c)\&(f) Calculated UMC ($\Delta\sigma_{xx}/\sigma_{xx}$). Black solid and blue dashed lines represent $\eta=1$ and 0.1 meV, respectively.}
\end{figure}

As the twist angle decreases, the band dispersion turns flat in the low-energy regime (see SM). Figure~\ref{fig3} shows the band structure and the UMC for 2D TBG with $\theta = 1.5 ^\circ$~\cite{note}. In both the $\mathbf{B || I}$ and $\mathbf{B \perp I}$ cases, the UMC ($\Delta \sigma_{xx} / \sigma_{xx}$) is larger in the flat band window (-30 to 30 meV in Fig. 3a) between the single-particle gaps, and one order of magnitude smaller in the regime of more dispersive band. For $\mathbf{B \parallel I}$, large UMC peaks up to 22\% exists at the same energies (about $\pm 10$ meV) as the DOS peaks, which are caused by the saddle points of flat bands. The UMC is also large near the band edge and diminishes quickly in higher energies for the dispersive bands. One can see that regions close to a van Hove singularity, i.e., saddle points and band edges, contribute substantially to the UMC. As shown in Fig. 3d, as the product $\eta L$ increases from zero, one can find that the UMC increases fast until it quickly reaches the peak value. For $L$ in the $\mathrm{\mu m}$ size, $\Delta \sigma_{xx} / \sigma_{xx}$ reaches the peak value for $\eta$ in the order of $0.1\,\mathrm{meV}$, which corresponds to a picoseconds lifetime. In experiment, $L$ is usually several $\mathrm{\mu m}$ and $\eta \sim 1\,\mathrm{meV}$ (e.g. in Ref.~\cite{Cao2018Mott}, the lifetime is $\sim 0.1\,\mathrm{ps}$).

\begin{figure}
    \centering
    \includegraphics[width=\linewidth]{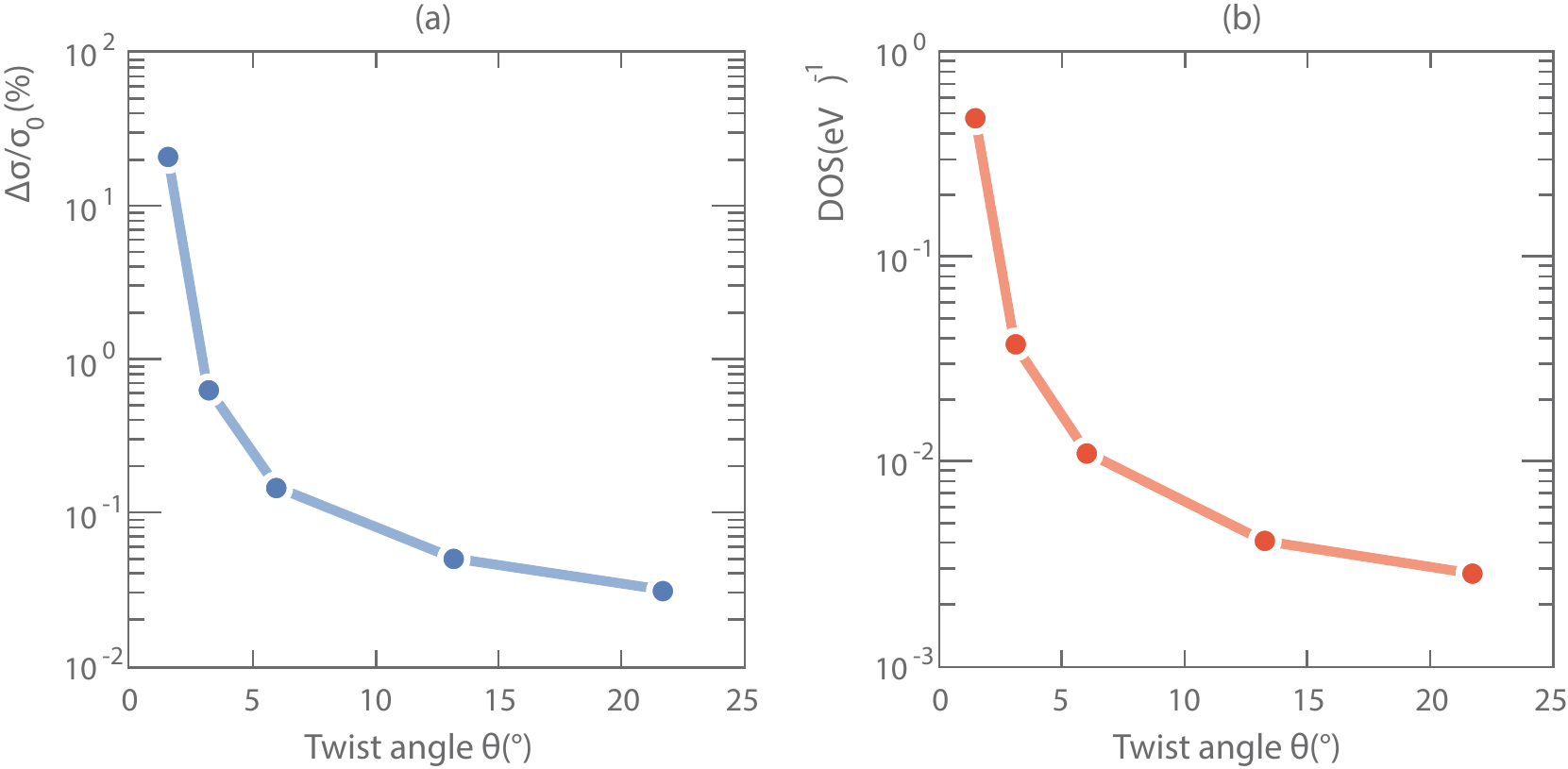}
    \caption{\label{fig4} \textbf{UMC and DOS.} The two-terminal (2T) UMC calculated by Eq. \eqref{Eq:UMR1} (a) and DOS (b) at peak position right below the charge neutral point for TBG at different angles. The UMC is approximately proportional to the DOS which agrees with Eq. \eqref{Eq:UMR2}.
    }
\end{figure}

The chirality-induced UMC/UMR increases quickly when the bands become flatter upon reducing $\theta$. Figure \ref{fig4}\textcolor{blue}{a} shows the $\theta$-dependence of the UMC peak value slightly below the charge neutral point (see the energy dependence in SM). The UMC peaks are proportional to corresponding DOS peaks (Fig. \ref{fig4}\textcolor{blue}{b}) at different $\theta$, well consistent with the observation from Eq.~\eqref{Eq:UMR2}.

For $\mathbf{B \perp I}$, we add a $\Delta_U = 50\,\mathrm{meV}$ potential difference to create a polarization $\mathbf{P}$ along the z-axis, leading to the $(\mathbf{B \times I)\cdot P}$ term for UMC/UMR. The UMC exhibits also large peaks in the flat band regime and at band edges of the dispersive bands. Two peaks below and above the charge neutral point are rather asymmetric compared to those in the $\mathbf{B || I}$ case, which is related to the asymmetry in the DOS (Fig. \ref{fig3}\textcolor{blue}{e}).

\section{Discussion}

\begin{figure}[t]
    \centering
    \includegraphics[width=0.9\linewidth]{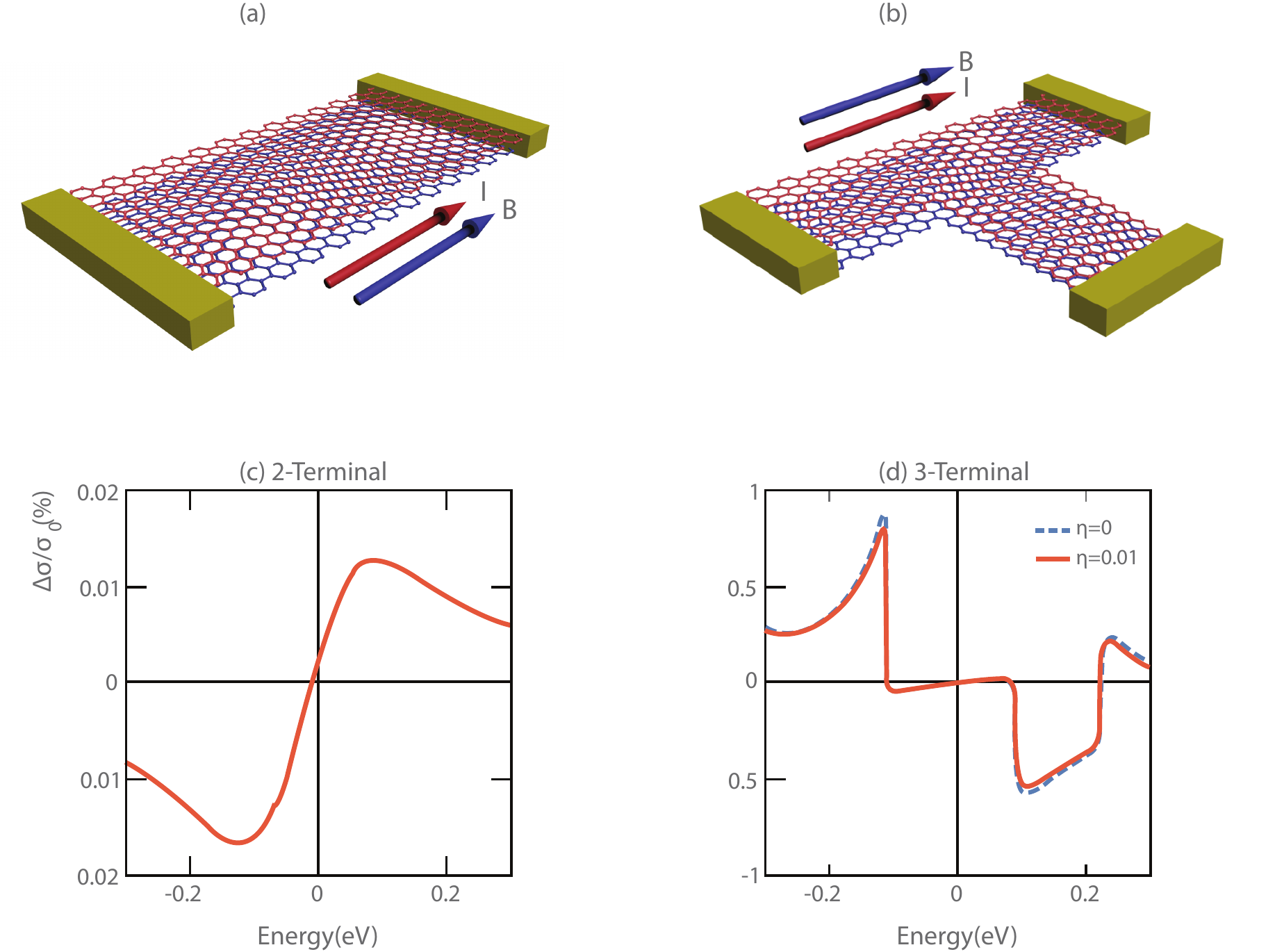}
    \caption{\label{fig5} \textbf{Three-terminal (3T) transport.} (a)-(b) Schematics of 2T and 3T devices. (c)-(d) Comparison of the 2T and 3T UMR calculated by NEGF method for $\mathbf{B||I}$ and B = 10 T. For the 2T device, a finite dephasing parameter $\eta=10$ meV is used; for the 3T one, both $\eta=0$ (red) and 10 meV (bule) are examined. }
\end{figure}

Although we discussed the UMR effect in a two-terminal setup, UMR can exist in multiple terminal devices, in which the UMR may even become stronger. In fact, devices with multiple terminals are commonly used in experiments (e.g., Refs.~\citenum{Cao2018Mott,Cao2018SC}). In two-terminal transport, the UMR requires a finite $\eta$ to act as a virtual lead which breaks charge conservation~\cite{Buttiker1986b,Liu2020chiral}. In cases with multiple terminals, charges can naturally leak to other leads. For example, Fig. \ref{fig5}{\textcolor{blue}{b}} shows a three-terminal device for TBG with $\theta=21.8^\circ$. The three-terminal UMC is less sensitive to $\eta$ and two orders of magnitude larger than the two-terminal result (Figs.~\ref{fig5}{\textcolor{blue}{c-d}}). We note that smaller angles show the same trend (see Fig. S5). Therefore, we expect the UMR to be even larger in multiple terminal experiments than in the above two-terminal results.



The current dependence of the resistance in Eq.~\eqref{Eq:EMCA} indicates a nonlinear phenomenon. In an experiment with an AC electric field, the UMR can be extracted from the second-harmonic resistance $R_{xx}^{(2\omega)}$ (e.g., Refs.~\citenum{Choe2019,Zhao2020WTe2,Yasuda2020QAHE}). For an angle $\phi$ between the in-plane field $\mathbf{B}$ and the current $\mathbf{I}$, the resistance has an elliptical dependence of the form $\alpha \cos{\phi} + \beta \sin{\phi}$, where $\alpha, \beta$ are parameters. This dependence is dissimilar to the case of AMR, which has the form of $\sin{2\phi}$. Therefore, while in the experiment both UMR and AMR may coexist, they can be disentangled by the periodicity in the angular dependence.

As discussed above, UMR is sensitive to the symmetry of the TBG. In the ideal $D_6$ symmetry case, UMR exists only for $\mathbf{B||I}$. In the $C_3$ case with interlayer potential difference, UMR appears also for $\mathbf{B \perp I}$ where $\mathbf{B}$ is still in-plane. In the lowest symmetry case $C_1$ with possible in-plane strain, UMR further allows for the out-of-plane $\mathbf{B}$ field, where, however, the ordinary $B^2$ term is usually overwhelming. 

The nonreciprocal transport was recently found to be strongly enhanced by superconductivity in chiral or noncentrosymmetric superconductors~\cite{Nagaosa2017,Iwasa2017,Lustikova2018}. It will be intriguing to investigate the UMR effect in the presence of superconductivity for TBG. Furthermore, in experiment magic-angle TBG can exhibit orbital ferromagnetism~\cite{Goldhaber2019,Young2019,Efetov2019,Wang2020,Li2020stm,Chen2020,Polshyn2020} due to the electron-electron interaction \cite{Po2018,Xie2020,Bultinck2020,Zhang2019,Liu2019,Wu2020,He2020}. We point that the orbital magnetism also leads to UMR where it plays the same role as the magnetic field. The orbital moment may modify the band structure much stronger than the magnetic field, inducing even larger UMR {if the two-fold out-of-plane rotational symmetry is further broken by lattice strain or substrate effects}. 

In summary, we report the giant UMR in twisted bilayer graphene as an orbital-derived phenomenon in the magnetic field. The UMR originates in the chiral atomic structure and gets enhanced by flat bands. We argue that the predicted UMR should be detectable in current experimental conditions. Besides quantum-transport calculations, we derived simple formulas (Eqs.~\ref{Eq2-G}-\ref{Eq2-R}) to monitor the UMR effect from the band structure, which is based on the direction-sensitive mean free path. Although the UMR is dependent on the dephasing and device geometry, it approaches a finite value in the strong scattering limit. This UMR limit is an intrinsic material quantity that depends on Fermi velocities of the band structure. {Thus, the UMR provides a direct probe to the flatness of the dispersion.}
In addition, our findings for TBG can be generalized to other chiral twisted bilayers and chiral crystals.

\section*{References and Notes}

\noindent \textbf{Acknowledgements --}
We thank the inspiring discussions with Yuval Oreg, Jiewen Xiao, Shahal Ilani, and Eli Zeldov. B.Y. acknowledges the financial support by the Willner Family Leadership Institute for the Weizmann Institute of Science, the Benoziyo Endowment Fund for the Advancement of Science,  Ruth and Herman Albert Scholars Program for New Scientists, the European Research Council (ERC Consolidator Grant No. 815869, ``NonlinearTopo'').

\noindent \textbf{Competing Interests:} The authors declare that they have no competing interests.

\noindent \textbf{Data and materials availability:} All data needed to evaluate the conclusions in the paper are present in the paper and/or the Supplementary Materials. Additional data related to this paper may be requested from the authors.

%

\end{document}